# Room temperature polarization in the ferrimagnetic $Ga_{2-x}Fe_xO_3$ ceramics

B. Kundys, F. Roulland, C. Lefèvre, C. Mény, A. Thomasson, N. Viart

*Institut de Physique et Chimie des Matériaux de Strasbourg (IPCMS), UMR 7504 CNRS-UdS, 67034 Strasbourg, France*

**Abstract**

The effect of the Fe/Ga ratio on the magnetic and electric properties of the multiferroic $Ga_{2-x}Fe_xO_3$ compound has been studied in order to determine the composition range exhibiting magnetic and electric orders coexistence and their critical temperatures. A magnetoelectric phase diagram, showing the evolution of both the Néel magnetic ordering temperature $T_N$ and the electric ordering temperature $T_c$, *versus* the iron content has been established for $0.9 \leq x \leq 1.4$. While the ferrimagnetic Néel temperature increases with the iron content, the electric ordering temperature shows an opposite trend. The electric polarization has been found to exist far above room temperature for the $x = 1.1$ composition which shows the highest observed electric ordering temperature of $T_c \approx 580\,K$. The $x = 1.3$ and 1.4 compounds are ferrimagnetic–electric relaxors with both properties coexisting at room temperature.

## 1. Introduction

The coexistence of magnetic and electric orders at the same temperature and pressure regions present high research interest due to the potential to design technologically important cross-functionalities in these materials. Such functionalities can include, but are not limited to, the electric field-controlled magnetization or magnetic field-controlled polarization.[1–3] There are however very few materials that present such properties at room temperature.[4,5] Until recently the only material showing unambiguously both ferroelectricity and magnetoelectricity at room temperature was $BiFeO_3$;[6–8] this explains why this material has been the focus point of most of the researches on multifunctional materials. Lately, convincing evidences of the coexistence of room temperature electric and magnetic orders and room temperature magnetoelectric effect have been observed in a few other compounds such as solid solution phases between lead iron based perovskites, and in particular the $PbFe_{0.5}Ta_{0.5}O_3$–$PbZr_{0.53}Ti_{0.47}O_3$ (PFT–PZT)[9,10] and $PbFe_{0.5}Nb_{0.5}O_3$–$PbZr_{0.53}Ti_{0.47}O_3$ (PFN–PZT)[11] systems, the complex hexaferrite $Sr_3Co_2Fe_{24}O_{41}$,[12] or even $\beta$-$NaFeO_2$.[13] The preparation of these compounds is however rather tricky and there is still room for improvement of their electric polarization, net magnetization and magnetoelectric effect.

The $Ga_{2-x}Fe_xO_3$ (GFO) compound, is known since the 60s to present interesting pyroelectric, ferrimagnetic and magnetoelectric properties near room temperature.[14] It crystallizes in the orthorhombic space group $Pc2_1n$, with $a = 0.87512 \pm 0.00008\,nm$, $b = 0.93993 \pm 0.00003\,nm$ and $c = 0.50806 \pm 0.00002\,nm$ (Fig. 1).[15] This structure is based on an ABAC double hexagonal close-packed stacking of oxygens, and strongly differs from the usual perovskite structure observed for most of the other multiferroic materials. The cations are distributed among four cationic sites Ga1, Ga2, Fe1, and Fe2.

Ga1 and Fe1 are antiferromagnetically coupled to Ga2 and Fe2. If the $Fe^{3+}$ cations only occupied the Fe1 and Fe2 sites, and the $Ga^{3+}$ only the Ga1 and Ga2 one, the compound would be strictly antiferromagnetic for $x = 1.0$. In fact a cationic site disorder, observed by Arima et al.[17] by neutron diffraction, allows the compound to exhibit a non-negligible net resulting magnetization (0.7 $\mu_B$/Fe for $x = 1.0$). The magnetic ordering temperature is relatively high in this family of compounds and is above room temperature for $x \geq 1.3$. Although $GaFeO_3$ was already reported to show magnetic field dependent polarization[14,17] the answer to the question whether it is electrically polar at room temperature and in zero magnetic field remains unclear. A phase transition to electrically polar state may be confirmed by a dipole reorientation induced maximum in the temperature variation of the dielectric permittivity. However the temperature variations of the dielectric properties reported so far for pure GFO compounds concern exclusively the $x = 1.0$ compound and the temperature dependence of the permittivity shows no maximum, indicating the absence of the transition to electrically polar state.[18–21] The only evidence of a maximum in the temperature variation of the dielectric permittivity was found for a Mn-doped GaFeO$_3$ sample.[22] Recently, a partially reversible polarization of *ca.* 0.3 µC/cm$^2$ was however evidenced in polycrystalline GaFeO$_3$ through pyroelectric measurements.[23] The reported ferroelectric Curie temperature ($T_{Cf}$) was below room temperature (*ca.* 100 K). A similar value of the polarization, *ca.* 0.2 µC/cm$^2$, but this time fully reversible and at room temperature, has been observed on thin films, both for the $Ga_{0.6}Fe_{1.4}O_3$:Mg[24] and GaFeO$_3$[25] compositions. It must be noted that a polarization of *ca.* 1 µC/cm$^2$ has also been measured on thin films of the isostructural compound $\varepsilon$-Fe$_2$O$_3$ at room temperature.[26] The awaited polarization in GaFeO$_3$ is however two orders of magnitude bigger than the measured ones, with a value of *ca.* 25 µC/cm$^2$, as evaluated by Stoeffler using a simple point charge model.

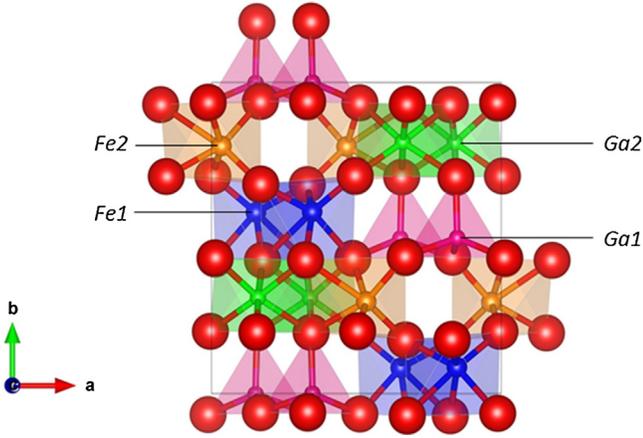

Fig. 1. Projection of the crystal structure of $Ga_{2-x}Fe_xO_3$ along the $c$ axis (space group $Pc2_1n$) produced using the VESTA[16] crystallographic software.

There is therefore a need to, first, clarify the electric behavior of GFO compounds and, then, study the influence of the iron content $x$ on the temperature of the prospective electric order. In this work, we attend to the electric characterization of a series of $Ga_{2-x}Fe_xO_3$ polycrystalline compounds ($0.9 \leq x \leq 1.4$) through both the study of the variation of the dielectric constant with temperature and pyroelectric current measurements.

## 2. Materials and methods

In our experiment the $Ga_{2-x}Fe_xO_3$ polycrystalline samples ($0.9 \leq x \leq 1.4$) were prepared *via* an optimized ceramic process already published elsewhere.[27] High purity commercial powders of $\alpha$-Fe$_2$O$_3$ (Prolabo >99%) and Ga$_2$O$_3$ (Alfa Aesar 99.999%) were first ball-milled in a Teflon jar in an optimized dispersive environment. The resulting slurries are subsequently dried and the powders obtained are mixed with an organic binder (Rhodoviol) to be uniaxially pressed under 60 bars into 15 mm diameter and 0.6 mm height disk shaped pellets using a hydraulic press. The green pellets were then sintered at the temperature necessary for the formation of the desired $Ga_{2-x}Fe_xO_3$, phase exempt of any parasitic phase, which depends upon the Fe content.[27] For the electric measurements, the temperature variation was performed in an Instec cryostat with a possibility to heat up to 873 K. Dielectric measurements were performed in vacuum at different frequencies using an Agilent LCR meter. The samples were first taken to the highest available temperature (*ca.* 600 K) and the dielectric constant was then measured upon cooling them down to 150 K. Pyroelectric current measurements were performed with a Keithley electrometer upon cooling the samples in a small electric field of 0.05 V/0.6 mm. The electric polarization was deduced from those pyroelectric measurements through a time integration method. Magnetic measurements were performed using a superconducting quantum interference device magnetometer (SQUID MPMS XL, Quantum Design).

## 3. Results and discussion

The temperature dependence of the dielectric permittivity was measured for various compounds of the $Ga_{2-x}Fe_xO_3$ family (with $0.9 \leq x \leq 1.4$). The results are shown in Fig. 2(a), and the evolution of the dielectric losses with the Fe content are presented in Fig. 2(b). A maximum in the temperature dependent permittivity is clearly seen for compositions above $x = 1.0$, indicating thus the existence of an electric polarization. The position of this maximum varies from 570 K for the $x = 1.1$ sample to about 400 K for the $x = 1.3$ sample. It is clear that both the electric ordering temperature ($T_c$) and dielectric losses of the GFO compounds strongly depend on the Fe/Ga ratio. The maximum in the dielectric permittivity curve is within an experimentally reachable range of temperatures for $x \geq 1.1$, but it strongly smoothens with increasing $x$.

The $Ga_{2-x}Fe_xO_3$, $x = 1.1$ compound shows the most important dielectric anomaly. The temperature position of the observed peak in the dielectric permittivity and dielectric loss is frequency independent, and allows determining a transition temperature of 582 K (Fig. 3(a)). The ratio between the slopes of the $1/\varepsilon'$ *versus* temperature curve above and below $T_c$ is larger than 2 (Fig. 3(b)), thus indicating a first-order transition to the paraelectric state above 582 K.

Pyroelectric current measurements also reveal an anomaly near this temperature for $x = 1.1$ (Fig. 4). The current integration with respect to time method gives a rather large polarization value of *ca.* 33 µC/cm$^2$, existing at room temperature. Although polarization loops have been measured in thin films of this material by different teams, including ours,[24,28] it has to be noted that our attempts to measure ferroelectric loops in the bulk compound, at different temperatures between 150 and 500 K, and in electric fields up to 60 kV/m, were unsuccessful. There is therefore no evidence of ferroelectricity of the material in its bulk form. Although the reported value of 33 µC/cm$^2$ has an only indicative character as it was induced by a specific electric field, it nevertheless, together with a peak in the dielectric permittivity, confirms the existence of an important electric polarization state in the sample, with a phase transition at 580 K. The $I(T)$ curve, measured under an applied electric field of 0.05 V/0.6 mm (Fig. 4) shows a zero current value in the temperature range of 160–450 K. This therefore excludes any artifact due to a dominant leakage contribution. At the electric ordering temperature, the small applied electric field allows orienting the polarization from random paraelectric state to polar state. It also has to be noted that the observed polarization value perfectly matches with the value of 25 μC/cm² calculated by Stoeffler[29] using first principle methods.

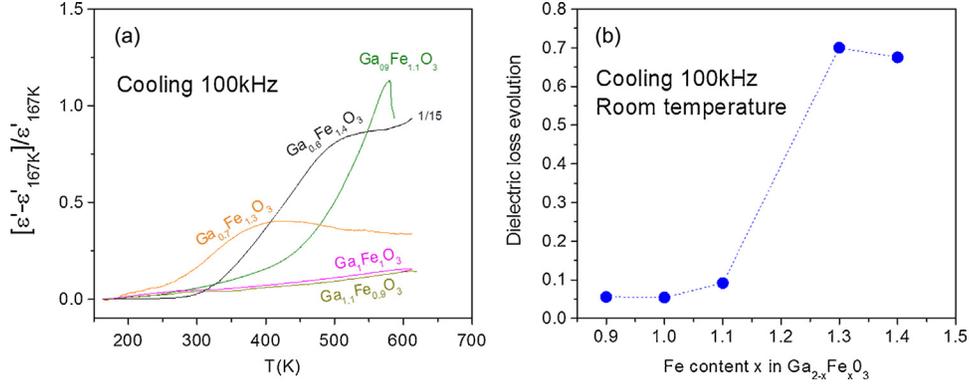

Fig. 2. (a) Temperature variation of the reduced dielectric permittivity measured at 100 kHz for various $Ga_{2-x}Fe_xO_3$ compounds ($0.9 \leq x \leq 1.4$) and (b) dielectric loss evolution with the Fe content in $Ga_{2-x}Fe_xO_3$ compounds at room temperature.

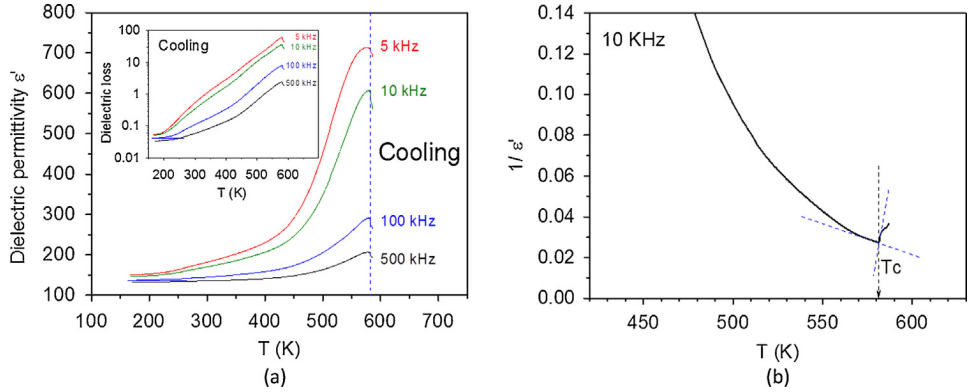

Fig. 3. (a) Frequency dispersion of the dielectric permittivity and dielectric loss (inset) as a function of temperature, (b) temperature dependence of $1/\varepsilon'$ below and above the electric ordering temperature $T_c$, for GFO $x = 1.1$.

For the $Ga_{2-x}Fe_xO_3$, $x = 1.3$ and 1.4 compounds, a relaxor-like behavior is observed, where the temperature for which the maximum permittivity is observed depends on the measurement frequency (Fig. 5).

Such a behavior is correlated with an important increase of the dielectric losses with the increasing iron content (Fig. 2(b)). The general trend in the $T_c$ evolution can be deduced by comparing the dielectric anomalies at one chosen frequency (Fig. 2(a)): it clearly decreases with increasing $x$.

The thermomagnetic curves of the samples were measured under an applied field of 75 Oe, high enough to increase the magnetic response while still keeping the samples well below magnetic saturation. The evolution of the Néel temperature of the samples *versus* their iron content is given in Fig. 6. As awaited,[17] the magnetic properties are strongly dependent on the iron content of the samples: the higher the iron content, the higher both the magnetization and the magnetic transition temperature. The most interesting results are ascribed to the two samples of highest iron contents ($x = 1.3$ and 1.4) as their $T_N$ values are higher than room temperature ($T_N = 330$ K for $x = 1.3$ and $T_N = 400$ K for $x = 1.4$). The $Ga_{2-x}Fe_xO_3$ system belongs to type 1 multiferroic compounds in which magnetic and electric orders occur at different temperatures.

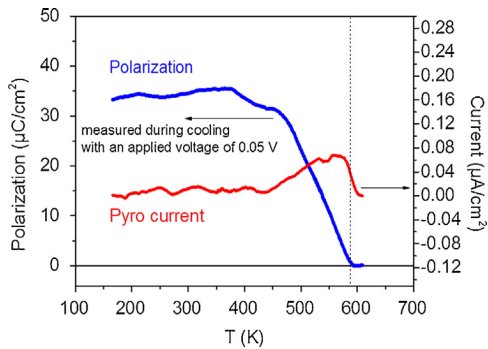

Fig. 4. Electric polarization (left) deduced from pyroelectric current (right) integration with respect to time, for GFO ($x = 1.1$).

4     *B. Kundys et al.*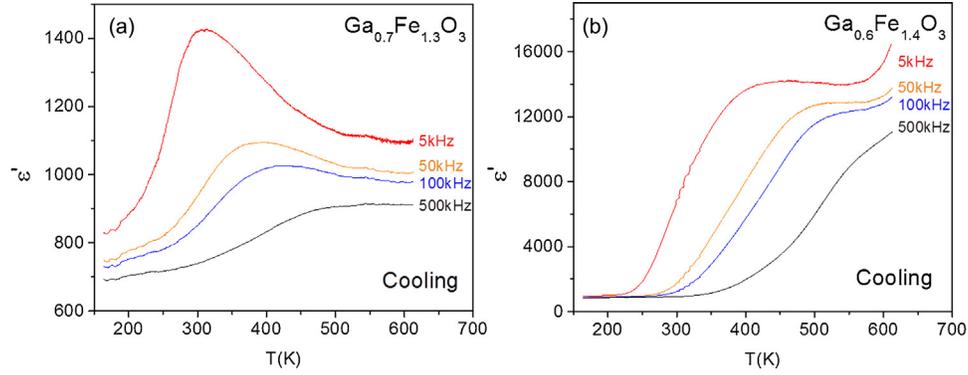

Fig. 5. Temperature variation of the dielectric permittivity for (a) $Ga_{0.7}Fe_{1.3}O_3$ and (b) $Ga_{0.6}Fe_{1.4}O_3$ at different frequencies.

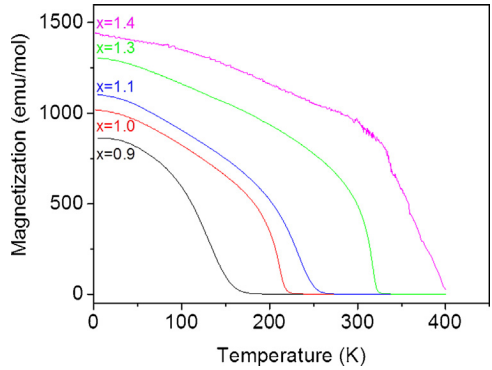

Fig. 6. M(T) dependences for the different GFO compounds ($0.9 \leq x \leq 1.4$) under an applied field of 75 Oe.

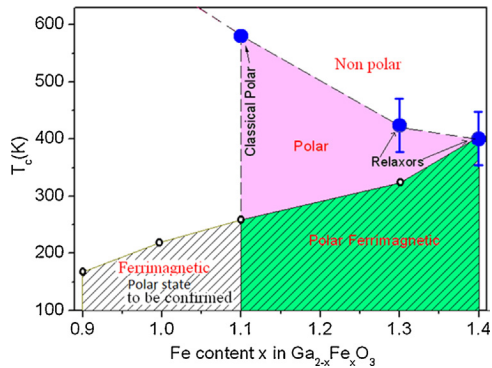

Fig. 7. Magnetoelectric phase diagram of $Ga_{2-x}Fe_xO_3$ compounds with $x$ between 0.9 and 1.4.

Fig. 7 finally summarizes the magnetoelectric phase diagram for $Ga_{2-x}Fe_xO_3$ compounds, showing the evolution of the Néel magnetic ordering temperature $T_N$, together with the electric ordering temperature $T_c$, *versus* the iron content.

## 4. Conclusion

In summary, the temperature variation (150 K < $T$ < 600 K) of the dielectric properties of $Ga_{2-x}Fe_xO_3$, $0.9 \leq x \leq 1.4$, compounds shows a maximum for the $x > 1$ compositions. The absence of previous documented studies for compositions other than $x = 1$ explains why this phenomenon had not been observed before. The composition $x = 1.1$ shows a polar state with an ordering temperature of about 580 K and a polarization of 33 µC/cm$^2$ very close to the value awaited from theoretical calculations. Compositions with $x \geq 1.3$ present a relaxor behavior, with high dielectric losses. The temperature at which the maximum of permittivity is observed decreases with increasing iron content. The dependence of the electric and magnetic properties of this system upon the Fe content is interestingly opposite, since $T_N$ increases with $x$. For samples with $x \geq 1.3$, the coexistence of both electric and magnetic polarizations in a wide temperature range including room temperature is possible. Further studies in this area should focus on the optimal $x$ values for which the dielectric relaxor behavior is transformed into a classical one with low dielectric dissipation, allowing a large magnetization and a large polarization near room temperature.

## Acknowledgments

This work was done with the financial support from the international ANR DFG Chemistry project GALIMEO #2011-INTB-1006-01. The authors are grateful to Dr. Ingrid CAÑERO INFANTE for fruitful discussions.

## Acknowledgments

This work was done with the financial support from the international ANR DFG Chemistry project GALIMEO #2011-INTB-1006-01. The authors are grateful to Dr. Ingrid CAÑERO INFANTE for fruitful discussions.